\documentstyle[aaspptwo,epsf]{article}

\begin{document}

\title {The Linear Sizes of Quasars and Radio Galaxies in the Unified Scheme}

\author{Gopal-Krishna and V.K.\ Kulkarni}

\affil{National Centre for Radio Astrophysics, Tata Institute of
Fundamental Research,\\ Poona University Campus, Post Bag.\ No.\ 3,
Pune 411007, India\\
e-mail: krishna(vasant)@gmrt.ernet.in}

\author{and Paul J.\ Wiita}

\affil{Department of Physics \& Astronomy, Georgia State
University, Atlanta, GA 30303-3083\\
e-mail: wiita@chara.gsu.edu}

\begin{abstract}
A key argument in favor of orientation based unification schemes is the
finding that among the most powerful 3CRR radio sources the 
(apparent) median linear size of quasars is smaller than that of
radio galaxies, which supports the idea that quasars
are a subset of radio galaxies, distinguished by being viewed at
smaller angles to the line of sight.
Recent measurements of radio sizes for a few other low frequency samples are,
however, not in accord with this trend, leading to the claim that orientation may not
be the main difference between radio galaxies and quasars.  We point out that
this ``inconsistency'' can be removed by making allowance for the temporal
evolution of sources in both size and luminosity, as inferred from 
independent observations.
This approach can also readily explain the other claimed ``major discrepancy''
with the unified scheme, namely, the 
difference between the radio luminosity--size correlations for quasars
and radio galaxies.

\end{abstract}

\vskip 1.0 true in
\centerline{\bf Accepted by ASTROPHYSICAL JOURNAL LETTERS on 13 March 1996}

\section{Introduction}

 According to a  widely discussed unified scheme for 
powerful F-R II extragalactic radio sources (with luminosities above $\sim 
10^{25.5}$
W Hz$^{-1}$ at 1 GHz, where we take $H_0 = 50$ km s$^{-1}$Mpc$^{-1}$ and $q_0 = 0.5$ throughout this {\it Letter}), narrow-line radio
galaxies (RGs) are identified as quasars (QSRs), or even blazars, whenever their 
principal axis happens to be oriented within a certain critical angle,
 ${\psi}$, from the line-of-sight (for recent reviews, see Antonucci 1993; 
Urry \& Padovani 1995; Gopal-Krishna 1995, 1996). In this model, the
parsec-scale nuclear core of such sources, consisting of a compact central 
engine ejecting relativistic jets of non-thermal continuum emission, and a
broad line region (BLR), is believed to be surrounded by a dusty torus.
In the case of RGs, the torus is believed to obscure the core 
in the visible through soft X-ray bands, so  that the BLR is not directly visible.
Strong
evidence for such tori comes from the detection of the BLR in the (scattered)
polarized light (e.g., Antonucci \& Miller 1985; Cimatti et al.\ 1993; Draper, Scarrott \&
Tadhunter 1993). Various evidences for the relativistic beaming aspect of 
this picture include: extremely rapid variability of blazars over all bands;
apparent superluminal motion in the bright radio cores; larger ratios of core
to total radio emission, $f_c$, in QSRs than in RGs; correlation of $f_c$
with the polarized optical nuclear continuum and its anti-correlations with
apparent radio linear size, the symmetry of core-to-lobe separations, and 
the equivalent width of the [O II]$\lambda$3727 line (Urry \& Padovani 1995;
Gopal-Krishna 1995, and references therein). The unified scheme is 
further supported by the analysis of the radio luminosity functions
of the postulated parent and aligned populations (Padovani \& Urry 1992).

One of the cornerstones of the orientation-based
paradigm for powerful radio sources has been the observation that in the
low-frequency 3CRR
sample (Laing, Riley, \& Longair 1983), where the axes of the sources
should be randomly oriented, 
 the median linear extent, $\ell$, of the extended radio emission from QSRs 
{\it appears} significantly smaller
   ($\stackrel{<}{\sim}$50\% for redshifts $z >$ 0.5 ) than that of RGs
(Barthel 1989).
However, the lack of such a behavior in the 
same sample at 
    $z <$ 0.5 (Kapahi 1990; Singal 1993a),
bolstered by similar trends reported recently for a few
   other low-frequency samples a few times deeper in flux density than the 3CRR, has evoked serious
   doubts about the unified scheme (Singal 1995, 1996; Kapahi et al.\ 1995).
Here we show that this apparently
   irrefutable inconsistency with the data can be resolved quantitatively
   by taking into account the available strong independent evidence for temporal evolution in both the sizes and luminosities of extragalactic double radio sources.

\section{Temporal and Luminosity Dependences of Radio Source Properties}

   Powerful radio sources are generally accepted to have rather limited
lifetimes; these are typically estimated to be 
    of the order of 10$^7$ yr, using the measured spatial
  gradients of spectral index across their radio lobes (e.g., Alexander
\& Leahy 1987; Liu, Pooley \& Riley 1992). 
Although any particular ``spectral age'' determination is uncertain,
and probably somewhat of an underestimate because of the complications
arising from field inhomogeneities (Siah \& Wiita 1990; Eilek 1996),
inverse Compton losses of
electron energy to the microwave background radiation (Wiita \& Gopal-Krishna 
1990), and
backflow from the jets into the radio lobes (Liu et al.\ 1992; Scheuer
1995), significant centimeter wave emission persisting for times in excess of
10$^8$yr is unlikely.  During this 
   active phase a double radio source grows in size through a hot
    circumgalactic medium at a speed $V$ which depends weakly on both the radio
    luminosity, $P$ ($V \propto P^{\alpha}, ~\alpha \ \simeq$ 0.3)
(Alexander \& Leahy 1987; Liu et al. 1992) and time ($V \propto
t^{\beta}, ~\beta \ \simeq  -$0.25 to 0) (Gopal-Krishna \& Wiita 1987,
1991; Fanti et al.\ 1996).
    Further, recent studies (Readhead 1995; Fanti et al.\ 1996),   
combining complete samples of powerful
    radio sources in different ranges of linear size, strongly suggest
    that the growth of a source from a few kpc to a few hundred kpc
    during the source lifetime is accompanied by typically
    an order-of-magnitude decrease
    in $P$ (largely due to expansion losses),
    after which the nuclear activity subsides rapidly (Rees 1994). Thus, we
    shall take $T$ = 10$^7$ yr as the characteristic e-folding time for the
 radio   luminosity decay, and  we approximate $P \propto$ exp($-t/T$), after a
short-lived initial brightening phase during which  the source turns on to a
 level
related to the power of its jets. While, admittedly, neither fully constrained
nor unique, this empirically supported characterization of luminosity evolution of individual sources
 appears to give a reasonable account of the temporal behavior of powerful
 extragalactic radio sources.

  A key element (Antonucci \& Miller 1985; Barthel 1989; Lawrence 1991)
of the unified picture is the critical angle $\psi$,
    which separates the sources classified as narrow-line objects (RGs) from those called
     broad-line objects (QSRs). Assuming that all powerful radio sources are intrinsically similar, this angle is readily determined from the
    fraction, $f_q$, of sources identified as QSRs 
    in a volume-limited sample of sources, if the sample is unbiased in orientation. 
 (Alternative possibilities, such as the assumption that a subset of RGs whose optical
spectra show only weak narrow-line emission would not exhibit broad lines from any
viewing direction [e.g., Laing et al.\ 1994], exist but will not be 
considered further here, as we are interested in examining whether the simplest
extensions of the unification scheme can be reconciled with the data.)
     In practice, such unbiased samples are selected at meter
    wavelengths, where the emission is dominated by radio lobes
    presumed to be radiating isotropically. Physically, $\psi$ is
identified with the half 
    angle of the polar openings of the dusty obscuring torus 
surrounding the active galactic nucleus (e.g.,
Antonucci \& Miller 1985; Antonucci 1993; Jaffe et al.\ 1993).  Although
    earlier investigations (Kapahi 1990; Lawrence 1991; Singal 1993a)
of the 3CRR sample suggested a systematic
    increase in $\psi$ with $P$ (or, $z$), this requirement was 
    found to be less compelling in some recent studies (Laing et al.\
1994; Saikia \& Kulkarni 1994) of the
    same 3CRR sample. Nonetheless, comparisons of $f_q$'s found 
    for the 3CRR and three other low-frequency complete samples
    which are selected at 408 MHz, have an expected $z$(median) $\sim 0.6$
 (Condon 1993), and are roughly an order-of-magnitude
    deeper  in flux density than the 3CRR sample, 
distinctly indicate a steady increase in $\psi$ with $P$ 
(Gopal-Krishna 1995; Singal 1995, 1996).
  Thus, beginning at large values of
$\sim$50$^\circ$--60$^\circ$ 
 for the most luminous radio sources (namely, the 3CRR sources at z $>$ 1),
 $\psi$ appears to shrink gradually
to 20$^\circ$ to 30$^\circ$ for sources in the samples selected at 408 MHz 
 which are likely to be two orders-of-magnitude less luminous, on average
(Gopal-Krishna 1995; Singal 1995, 1996). 
Note that we consider the dependence of $\psi$ to be primarily on (initial)
 $P$ instead of $z$, firstly because it is expected on theoretical grounds 
(see below), and secondly because such a dependence has also been inferred
for optically selected active galactic nuclei (Falcke,
Gopal-Krishna, \& Biermann 1995, and references therein). 

One likely theoretical  
explanation for the variation in torus opening angle involves the
action of  energetic nuclear photons on the
dust particles within the torus, so that more powerful central engines
may be expected to have larger values of $\psi$ (Netzer \& Laor 1993;
K\"onigl \& Kartje 1994). Note that 
for a given source, $\psi$ need not decrease monotonically with time, as the
secular decay of its radio luminosity (see above) can be expected to arise 
primarily from expansion losses (Scheuer 1974; Gopal-Krishna \& Wiita 1991),
rather than from a decline in the intrinsic jet power. The diminishing of 
the radio lobe pressure due to the expansion
would also lead to a lowering of the optical output of the narrow-line
emitting filaments within and around the lobe, thus explaining the
correlation of the radio luminosity $P$ with the narrow [O III] line emission,
established by Baum \& Heckman (1989) and Rawlings \& Saunders (1991). The
possibility exists, on the other hand, that $\psi$ may increase with time, due
to a steady rise in the mass of the central engine.   The greater gravitational dominance of a more massive black hole over the surrounding stellar core is expected to 
push outward the inner edge of the molecular torus, due to a shift in the 
region where reduced differential rotation allows a piling-up of the 
accreted gas (Yi, Field, \& Blackman 1994). Such a circumstance would
facilitate our explanation
for the apparent excess of the QSR radio sizes vis-\`a-vis RGs (see below).

\section{The Model}

    In light of the above observations and discussion, we adopt the
following simple relations
    to broadly represent the actual dependences for radio luminosity,
critical angle, and physical (deprojected) size, respectively:

\begin{equation}
P(t) = P_0 {\rm exp}(-t/T), ~~~{\rm where} ~T=10^7{\rm yr};  
\end{equation}
\begin{equation}
\psi(P_0) =  a~{\rm log}_{10}(P_0/10^{26}{\rm W Hz}^{-1}),
\end{equation}
where $0.2~{\rm radian}~ \le \psi \le \psi_{max} \simeq 1.0~{\rm  radian}$;
\begin{equation}
L(P_0,t) \propto P_0^{b_1} t^{b_2}; 
\end{equation}
\noindent where $~a, ~b_1$ and $b_2$ are constants (input parameters to
the model) whose characteristic values can be approximated from the
empirical estimates for $f_q$, $\alpha$, and $\beta$ mentioned above.

We can now compute the RG-to-QSR  median linear size ratio, $R$,
for a temporally evolving population of sources.
Consider a radio luminosity, $P$, which is below the
    value for the most powerful radio sources created ($P_{max}$). At any
    particular time, the sources {\it observed} at the luminosity level $P$
    would include not just the young (hence small) sources freshly created with
    that level of luminosity, but also the aging sources that were born 
    with higher luminosities in the past, but have since then faded down to
    $P$, and concurrently expanded to larger sizes. Since these older
  expanded radio sources with larger $\ell$ would have a higher QSR
fraction ($f_q$) arising from
    their higher initial $P_0$ (and correspondingly larger $\psi$), the
    median radio size of QSRs found in intermediate luminosity
    samples with $P$ substantially below $P_{max}$ may well approach, or even exceed,
 that of RGs. This could possibly be 
   the explanation for the reported radio size ``anomaly''
    mentioned above (Gopal-Krishna 1995), which is claimed (Singal
1993a, 1995; Kapahi et al.\ 1995) to rule out the
    orientation based unified scheme.

    To explore this proposal quantitatively, let us assume that 
    a certain number of sources is
    ``injected'' continuously at a fixed rate (arbitrarily set to 100 per
unit time) within a unit volume of space, 
    with radio luminosities distributed as
\begin{equation}
N(P_{0}) \propto P_0^{-\Gamma}, ~~~{\rm for} ~P_0 ~{\rm up~to} ~P_{max}. 
\end{equation}

Our simple model 
    parameterized by Eqns.\ (1--4) can now be used to numerically predict
    the quasar fraction, $f_q$ at a given $P$, as well as the corresponding 
    ratio, $R$, of the 
    median projected radio sizes of RGs and QSRs.  To do this we
    first need to derive the distribution function of ages of sources
of a given
current luminosity
    $P$.  To do so we start from the full distribution function of
$P$ and $t$, which is
\begin{equation}
\rho(P,t) = K P^{-\Gamma}{\rm exp}[-(\Gamma - 1)t/T].
\end{equation}

>From this we derive the (normalized) distribution function of $t$ for a 
given $P$ to be
\begin{equation}
\Phi(t\vert P = P) = { (\Gamma - 1)T^{-1}{\rm exp}[-(\Gamma - 1)t/T] \over
[1 - (P/P_{0,max})^{\Gamma - 1}]}.
\end{equation}
 We then generate a large number of sources, the ages of which
follow this distribution.  For each source we take its 
age, $t_i$, and its orientation angle from the line-of-sight, 
$\theta_i$ (randomly
selected from $0$ to $\pi/2$), yielding its $P_{0,i}$ (Eqn.\ 1)
and $\psi_i$ (Eqn.\ 2).  If $\theta_i < \psi_i$ then the object is
counted as a quasar, otherwise it is counted as a galaxy.  We then
obtain its linear size, $L$ (Eqn.\ 3), and projected linear size, $\ell$.  By
repeating this procedure many times, we compute the median projected linear
size for RGs and for QSRs (yielding $R$), as well as $f_q$, for the
adopted value of $P$.

    In Fig.\ 1 our simple model is confronted with the observations by
plotting the combinations of $f_q$  and $R$    
computed as above for different values of $P$, taking some characteristic values for the input parameters
based on the above discussion and using additional observations
to set $\Gamma \approx 2$ in Eqn.\ 6 (see Dunlop \& Peacock 1990).
 In Eqn.\ 4, $P_{max}$,  has been set equal
 to 10$^{29}$W Hz$^{-1}$ at $1~$GHz, characteristic of the most luminous
 radio sources.
The runs of the computed values of $f_q$  and  $R$ are shown by
three curves, each covering 
     a two orders-of-magnitude range in $P$.
      Approximately the same range is spanned 
by the data points,   which
are reproduced from Singal (1995), and  correspond 
    to subsets of the 3CRR, 1-Jy and B3 samples. 
Note that although redshifts
    for the B3 subset are not fully known, its median flux density of
    $\sim$1 Jy at 408 MHz suggests (Condon 1993) a $z$(median) $\sim$0.6,
 making it roughly two orders-of-magnitude less luminous, on average,
than the most luminous 3CRR subset ($z \ge 1$) shown by the rightmost
data point in Fig.\ 1. 
 We further note that (in the absence of complete redshift 
information) although the
 size ratios given in Singal (1995; also, Fig.\ 1) actually refer to angular
sizes, they are not likely to differ significantly from the linear size
ratios, as most of the sources involved lie at $z \ge 0.5$ where the
linear-to-angular conversion is relatively insensitive to $z$ (see also
Table 1 in Singal 1993a).  Moreover, a similar drop in the value of
$R$ to around unity  near $S_{408} \sim$1 Jy has been inferred (Kapahi et al.\ 1995)
from the linear sizes of the Molonglo QSRs and 3CRR RGs.

 As highlighted by Singal (1995, 1996), the data points 
provide a poor match to the prediction of the
 orientation-based unified scheme (shown by the curve ``U'' in
 Fig.\ 1) for any single value of $\psi$; even if $\psi$ were allowed
to vary, the predicted trend runs counter to the disposition  of the
data points. It is seen, however, that once an allowance is made for the
(inevitable) temporal evolution broadly characterized by our simple, empirically
 supported model, the predicted $f_q$--$R$ profiles for the
 unified scheme are found to match the data points quite well (Fig.\ 1).

An interesting implication of the present model is that except in the
meter wavelength samples selected at the highest luminosities, quasars
would be systematically {\it older} (and physically larger) than RGs.  This is
opposite to the situation envisioned in some alternative unification schemes
(e.g., Hutchings, Price, \& Gower 1988).

Finally, a cautionary remark, motivated by our implicit assumption that broad
and narrow-line objects can be distinguished at all luminosities and redshifts.
This assumption may not be valid at lower luminosities where any broad component
underlying a narrow spectral feature may be increasingly hard to detect,
leading to an underestimate of $f_q$ (and, therefore, $\psi$). The significance of
this potential bias cannot be quantified using the currently available data, though it
is unlikely to be a major effect for the relatively luminous objects being
considered here.

\section{Conclusions}

We emphasize that the temporal evolution incorporated here
into the orientational unified scheme is exceedingly simple and
expressed in terms of just a few parameters, all of which are
constrained by observations.  Since this evolutionary
scenario is inspired by observations and elementary theoretical
considerations, the neglect of this factor in the past attempts to verify
the unified scheme was a major shortcoming. Consequently, the claimed ``mismatch''
with the radio size data (Singal 1993a, 1995, 1996; Kapahi et al.\ 1995)
should not mandate dismissal of the paramount role assigned to
orientation effects in the unified scheme.

The present explanation for the decreasing RG-to-QSR size ratio,
$R$, towards lower radio luminosities also provides an explanation
for an {\it equivalent} observational result according to which
QSRs and RGs exhibit different $\ell$--$P$ correlations. Several authors
have pointed out that the empirically derived positive $\ell$--$P$
dependence for RGs ($\ell \propto P^x$, $x \simeq +0.3$), contrasts
with practically no $\ell$--$P$ dependence found for QSRs ($x \simeq 0$),
and have argued that this ``puzzling'' result effectively rules out
the unified scheme 
(Chyzy \& Zieba 1993; Singal 1993a,b, 1995, 1996; Kapahi et
al.\ 1995). However, this difference too, 
can be readily understood in our picture.  The values of $x$
 corresponding to the three
 computed model curves shown in Fig.\ 1 are  $+0.24 \pm 0.07$ for RGs
 and $0.09 \pm 0.08$   
    for QSRs, as deduced from the linear fits to the computed median sizes 
at different  radio luminosities for the three sets of input parameters;
for each fit the slope $x$ is found to be greater for RGs than for QSRs. As a
 refinement to the present analysis, the use of full linear size distributions
(rather than the median values) would be a useful step.	

In summary, we conclude that the linear size measurements of radio
galaxies and quasars
    reported up to this point do not  violate the basic tenet of the 
    unified scheme, viz, that radio galaxies are oriented closer to
the sky plane than are quasars.

\acknowledgments
   We thank the anonymous referee for suggestions which led to improvements in the presentation.  This work was supported in part by NSF grant AST-9102106, NASA grant
NAG 5-3098, and Chancellor's Initiative Funds at Georgia State University.



\begin{figure}
\plotone{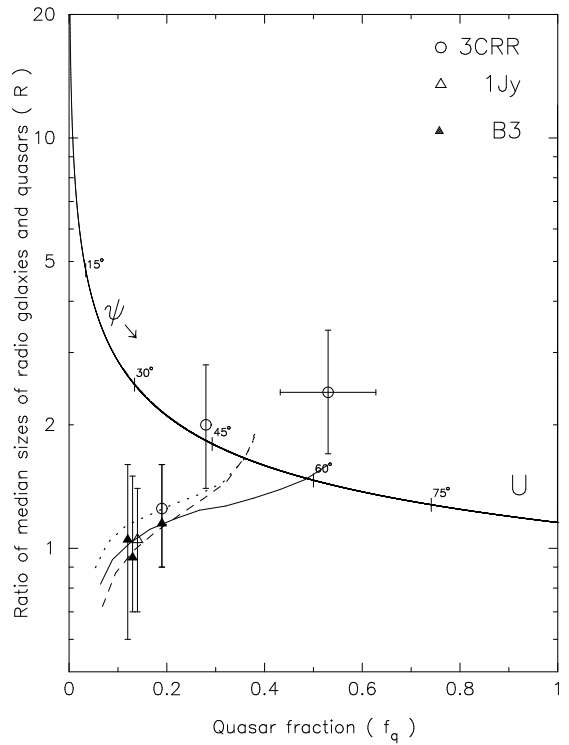}

\caption{A plot of $R$ {\it versus} $f_{q}$ for the subsets
of three flux-limited samples selected at meter-wavelengths. The data shown
with error bars are
reproduced from Singal (1995). The thin curve ``U" depicts the prediction of the
unified scheme for such orientation
bias-free samples, over a range in 
the torus half-angle $\psi$. The three other  curves represent 
our  predictions of the unified scheme incorporating  temporal
evolution, as described in the text. Each of the curves covers
 a two orders-of-magnitude
range in radio luminosity, increasing to the right. The input parameters
($\Gamma, a, b_1, \psi_{max}$) are (1.8, 0.38, 0.18, 1.1 rad) for the 
solid curve, (1.8, 0.40, 0.25, 0.9 rad) for the dashed curve, and 
(2.0, 0.40, 0.30, 0.9 rad) for the dotted curve.
The other input
parameters were set (following the empirical
results mentioned in the text) to:  $b_2 =$ 0.8 and $\psi_{min}$ = 0.2 rad.
Note that a horizontal error bar is also  shown for the  rightmost
data point; it  represents  a typical 1$\sigma$ error of these 3CRR
data points, each of which 
is based on a relatively  small set of sources.} 
\end{figure}

\end {document}